\documentclass{aastex}
\usepackage{emulateapj5,epsf,apjfonts}

\slugcomment{to appear in ApJ}

\shorttitle{Abundance gradients and the role of SNe in M87}
\shortauthors{Gastaldello \& Molendi}

\begin{document}
   
\title{Abundance gradients and the role of SNe in M87}


\author{Fabio Gastaldello}
\affil{IASF - CNR, via Bassini 15, I-20133 Milano, Italy}
\affil{Universit\`a di Milano Bicocca, Dip. di Fisica, P.za della Scienza 3 I-20133 Milano, Italy}

\email{gasta@ifctr.mi.cnr.it}

\and

\author{Silvano Molendi}
\affil{IASF - CNR, via Bassini 15, I-20133 Milano, Italy}
\email{silvano@ifctr.mi.cnr.it}


\begin{abstract}
We make a detailed measurement of the metal abundance profiles and metal 
abundance ratios of the inner core of M87/Virgo observed by \emph{XMM-Newton} during
the PV phase. We use multi temperature models for the inner regions and we
compare the plasma codes APEC and MEKAL. We confirm the strong heavy elements 
gradient previously found by \emph{ASCA} and \emph{BeppoSAX}, but also find a significant 
increase in light elements, in particular O. This fact together with the 
constant O/Fe ratio  in the inner 9 arcmin indicates an enhancement 
of contribution in the core of the cluster not only by SNIa but also by SNII. 

\end{abstract}



\section{Introduction}

The X-ray emitting hot intra-cluster medium (ICM) of clusters of galaxies is
known to contain a large amount of metals: for rich clusters between red-shift 
0.3-0.4 and the present day the observed metallicity is about 1/3 the solar
value \citep{mush97,fuka98,allen98,ceca00,ettori01}, 
suggesting that a significant fraction of the ICM has been processed
into stars already at intermediate red-shifts.

While the origin of the metals observed in the ICM is clear (they are produced 
by supernovae), less clear is the transfer mechanism of
 these metals to the ICM. The main mechanisms that have been 
proposed for the metal enrichment in clusters are: enrichment of gas during the
formation of the proto-cluster \citep{kauff98}; 
ram pressure stripping of metal enriched gas from cluster galaxies 
\citep{gunn72,toni01}; stellar winds AGN- or 
SN-induced in Early-type galaxies \citep{matteucci88,renzini97}.

Spatially resolved abundance measurement in galaxy clusters are of great 
importance because they can be used to measure the precise amounts of metals 
in the ICM and to constrain the origin of metals both spatially and in terms
of the different contributions of the two different type of SNe (SNII and SNIa)
as a function of the position in the cluster. The first two satellites 
able to perform spatially resolved spectroscopy, \emph{ASCA} and 
\emph{BeppoSAX} have revealed abundance gradients in cD clusters 
\citep{dupke00b,grandi01},
in particular M87/Virgo \citep{matsumoto96,guainazzi00}, 
and variations in Si/Fe 
within a cluster \citep{fino00} and among clusters 
\citep{fuka98}. Since the SNIa products are iron enriched, while the 
SNII products are rich in $\alpha$ elements, such as O, Ne, Mg and Si, the
variations in Si/Fe suggest that the metals in the ICM have been produced by
a mix of the two types of SNe.
The exact amount of these mix still remain controversial: 
\citet{mush96} and \citet{mush97} showed a 
dominance of SNII ejecta, while other works on \emph{ASCA} data 
\citep{ishimaru97,fuka98,fino00,dupke00a}, still indicating a predominance of 
SNII enrichment
at large radii in clusters, do not exclude that as much as 50\% 
of the iron in clusters come from SNIa ejecta in the inner part of clusters.  

M87 is the cD galaxy of the nearest cluster and its high flux and close 
location allows, using the unprecedented combination of spectral and 
spatial resolution and high throughput of the EPIC experiment on board 
\emph{XMM-Newton}, a detailed study of the ICM abundance down to the scale of the kpc.
Throughout this paper, we assume $H_{0}\,=\, 50\, \rm{km\, s^{-1} Mpc^{-1}}$, 
$q_{0} = 0.5$ and at the distance of M87 $1^{\prime}$ corresponds to 5 kpc.

Recently, when we were finishing the writing of the paper, we have learned of 
the \citet{fino01} analysis of the same data. Our work is 
complementary with theirs in the sense that we use 10 annular bins instead 
of 2, fully
exploiting the quality of the \emph{XMM} data and taking into account the multi 
temperature appearance of spectra in the inner regions with adequate 
spectral modeling. 

The outline of the paper is as follows. In section 2 we give information about
the \emph{XMM} observation and on the data preparation. In section 3 we suggest
the use of a unique set of ``standard'' abundances. In section 4 we 
describe our
spectral modeling. In section 5 we present spatially resolved measurements
of metal abundances and abundance ratios and in section 6 we discuss our 
results. A summary of our conclusions is given in section 7.

\section {Observation and Data Preparation}
M87/Virgo was observed with \emph{XMM-Newton} \citep{jansen01} during the PV 
phase with the MOS detector in Full Frame Mode for an effective exposure
 time of about 39 ks. Details on the observation have been published in 
\citet{boh01} and \citet{belsole01}.
We have obtained calibrated event files for the MOS1 and MOS2 cameras 
with SASv5.0. Data were manually screened to remove any remaining bright 
pixels or hot column. 
Periods in which the background is increased by soft proton flares have 
been excluded using an intensity filter: we rejected all events accumulated 
when the count rates exceeds 15 cts/100s in the [10 -- 12] keV band for the 
two MOS cameras.

We have accumulated spectra in 10 concentric annular regions centered on the 
emission peak extending our analysis out to 14 arcmin from the emission peak, 
thus exploiting the entire \emph{XMM} field of view. We have removed point 
sources and the substructures which are clearly visible from the X-ray image
\citep{belsole01} except in the innermost region, where we have kept
the nucleus and knot A, because on angular scales so small it is not possible 
to exclude completely their emission. We prefer to fit the spectrum of this
region with a model which includes a power law component to fit the two point 
like sources. We include only one power law component due to the similarity
of the two sources spectra \citep{boh01}.
The bounding radii are 0$^{\prime}$-0.5$^{\prime}$, 
0.5$^{\prime}$-1$^{\prime}$, 1$^{\prime}$-2$^{\prime}$, 
2$^{\prime}$-3$^{\prime}$, 
3$^{\prime}$-4$^{\prime}$, 
4$^{\prime}$-5$^{\prime}$, 5$^{\prime}$-7$^{\prime}$, 
7$^{\prime}$-9$^{\prime}$, 9$^{\prime}$-11$^{\prime}$ and 
11$^{\prime}$-14$^{\prime}$. 
The analysis of the 4 central 
regions within 3 arcmin was already discussed in \citet{molegasta01}.

Spectra have been accumulated for MOS1 and MOS2 independently. The Lockman 
Hole observations have been used for the background. Background spectra have 
been accumulated from the same detector regions as the source spectra.

The vignetting correction has been applied to the spectra rather than to the
effective area, as is customary in the analysis of EPIC data 
\citep{arnaud01}. Spectral fits were performed in the 
0.5-4.0 keV band. Data below 0.5 keV were excluded to avoid 
residual calibration problems in the MOS response matrices at soft energies.
Data above 4 keV were excluded because of substantial contamination 
of the spectra by hotter gas emitting further out in the cluster, on the
same line of sight.

As discussed in \citet{molendi01} there are cross-calibration uncertainties 
between
the spectral response of the two EPIC instruments, MOS and PN. In particular 
for what concern the soft energy band (0.5-1.0 keV) fitting six 
extra-galactic spectra for
which no excess absorption is expected, MOS recovered the $\rm{N_{H}}$ 
galactic value, while PN gives smaller $\rm{N_{H}}$ by 
$1-2\times10^{20}\rm{cm^{-2}}$. Thus we think that at the moment the 
MOS results
are more reliable than the PN ones in this energy band, which is crucial 
for the O abundance measure. For this
reason and for the better spectral resolutions of MOS, which is again important
 in deriving the O abundance, we limit our analysis to MOS data. 

\section{Solar Abundances}    

The elemental abundances of astrophysical objects are usually expressed by
 the relative values to the solar abundances. The so-called solar abundances
can be either ``meteoritic'' or ``photospheric''.

This distinction between ``meteoritic'' and ``photospheric'' solar abundances 
was made in the review by \citet{anders89}. Significant discrepancies
exist between the two sets of abundances quoted in that paper, particularly 
for iron and this has caused in the past some controversy in the discussion of 
the results of cluster abundances \citep{ishimaru97,gibson97}.
However recent photospheric models of the sun indicate that photospheric 
and meteoritic abundances agree perfectly and the community has converged 
toward a ``standard solar composition'' \citep{grevesse98}, with 
suggestions to the astrophysical community to accept this new 
state of the art \citep{brighenti99}.
For the above reason, in this paper we shall adopt the \citet{grevesse98}
 values.
Since the solar abundance table used by default in XSPEC is based on
photospheric values of \citet{anders89}, we have switched to a 
table taken from the data by \citet{grevesse98}
 by means of the XSPEC command ABUND.
In general a simple scaling allows to switch from one set of abundances to 
the other.

\section{Spectral modeling and plasma codes}

All spectral fitting has been performed using version 11.0.1 of the 
XSPEC package. 

All models discussed below include a multiplicative component
to account for the galactic absorption on the line of sight of M87.
The column density is always fixed at a value of 
$1.8\times 10^{20}\,\rm{cm^{-2}}$, which is derived from 21cm measurements 
\citep{lieu96}.
Leaving $\rm{N_{H}}$ to freely vary does not improve the fit and does 
not affect the measure of the oxygen abundance, which could have been 
the more sensitive to the presence of excess absorption. 
The $\rm{N_{H}}$ value 
obtained is consistent within the errors with the 21cm value.

The temperature profile for M87 \citep{boh01} shows a small 
gradient for radii larger than $\sim 2$ arcmin and a rapid decrease for 
smaller radii. Moreover, as pointed out in \citet{molepizzo01}  all 
spectra at radii larger than 2 arcmin are
characterized by being substantially isothermal (although 
the spectra of the regions between 
2 and 7 arcmin are multi temperature spectra with a narrow temperature 
range rather than single temperature spectra), while at radii smaller 
than 2 arcmin
we need models which can reproduce the broad temperature distribution of the
inner regions.

We therefore apply to the central regions (inside 3 arcmin) 
three different spectral models.

A two temperature model 
(vmekal + vmekal in XSPEC and model II in \citet{molegasta01} using
the plasma code MEKAL \citep{mewe85,liedahl95}. This
model has 15 free parameters: the temperature and the normalization of the
two components and the abundance of O, Ne, Na, Mg, Al, Si, S, Ar, Ca, Fe
 and Ni, all expressed in solar units. The metal abundance of each 
element of the second thermal component is bound to be equal to the same 
parameter of the first thermal component. This model is used (e.g. 
\citet{makishima01} and refs. therein) as an alternative to cooling-flow 
models in fitting
the central regions of galaxy clusters.

A ``fake multi-phase'' model (vmekal + vmcflow in XSPEC and model III 
in \citet{molegasta01}. This model has 15 free parameters, as
the two temperature model, because the maximum temperature 
$T_{max}$ is tied to the vmekal component temperature. 
As indicated in recent papers 
\citep{molepizzo01,molegasta01} this model is used to
describe a scenario different from a multi-phase gas, for which it was 
written for: the gas is all at one temperature and the multi-phase 
appearance of the spectrum comes from projection of emission from many 
different physical radii. A more correct description will be given by a real
deprojection of the spectrum (Pizzolato et al., in preparation).
  
The third model is the analogue of the vmekal two temperature model using
the plasma code APEC \citep{smith01}. This model has 14 free parameters,
 one less than the corresponding model using vmekal because APEC misses the
Na parameter. 

We can't adopt an APEC analogue of the fake multi-phase model because the 
cooling flow model calculating its emission using APEC is still 
under development. Given the substantial agreement between 2T and fake 
multi-phase model \citep{molegasta01}, we can regard the 2T APEC 
results as indicative also for a fake multi-phase model.

For the spectrum accumulated in the innermost region we included also a power
law component to model the emission of the nucleus and of knot A.

For the outer regions (from 3 arcmin outwards) we apply single temperature
models: vmekal using the MEKAL code, with 13 free parameters and vapec using
the APEC code, with 12 free parameters.

As pointed out by the authors of the new code, cross-checking is very 
important, since each plasma emission code requires choosing from a large 
overlapping but incomplete set of atomic data and the results obtained by using
independent models allows critical comparison and evaluation of errors in
the code and in the atomic database.
 
\section{Results}

\subsection{Abundance measurements and modeling concerns}

The X-ray emission in cluster of galaxies originates from the hot gas 
permeating the cluster potential well. The continuum emission is dominated by
thermal bremsstrahlung, which is proportional to the square of the gas density
times the cooling function. From the shape and the normalization of the 
spectrum we derive the gas temperature and density. In addition the X-ray
spectra of clusters of galaxies are rich in emission lines due to K-shell
transitions from O, Ne, Mg, Si, S, 
Ar and Ca and K- and
L-shell transitions from Fe and Ni, from which we can measure the
relative abundance of a given element.

\begin{figure}
\plotone{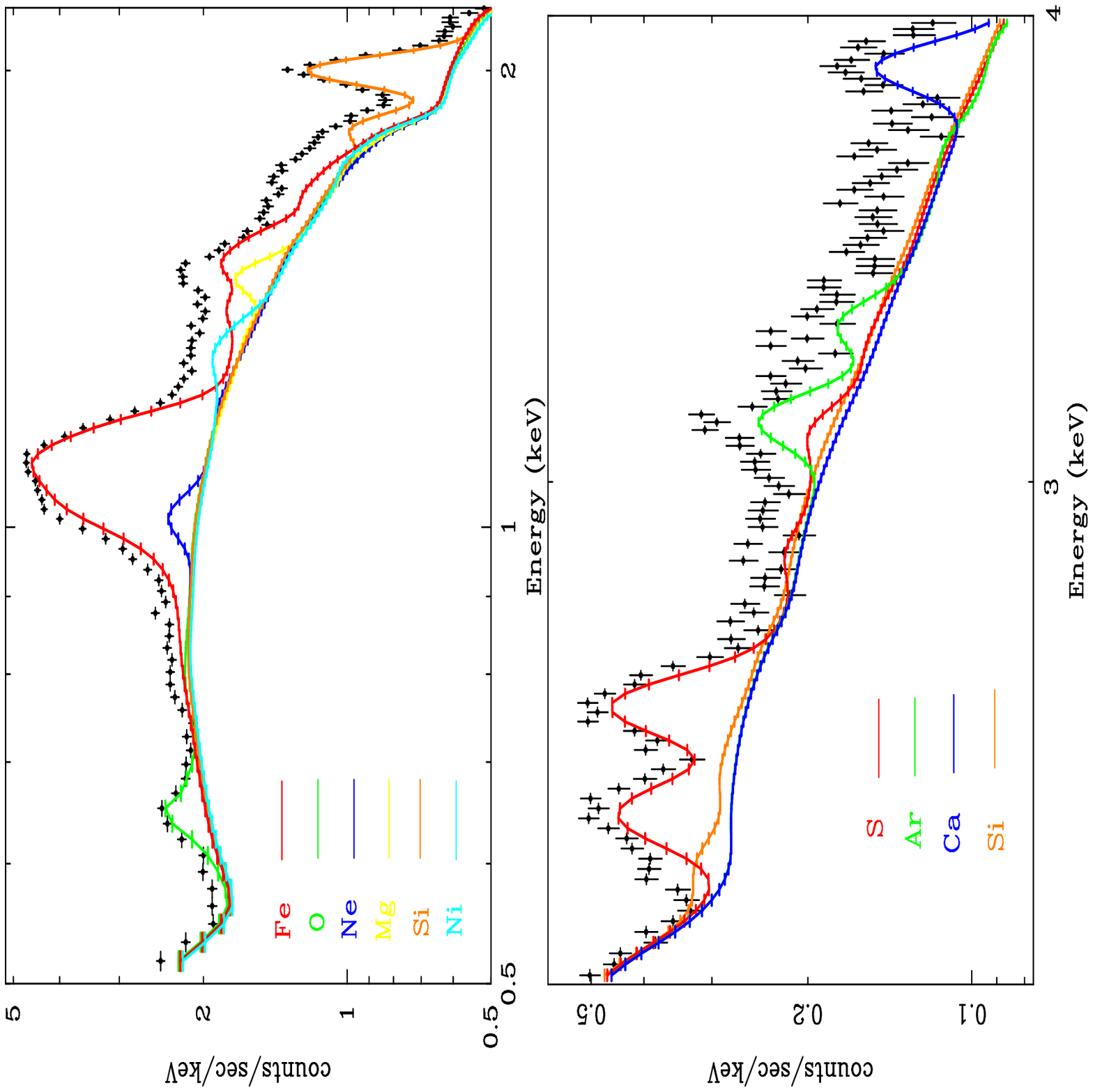}
\epsscale{0.9}
\vskip -3.5truecm
\caption
{Data of the 3$^{\prime}$-4$^{\prime}$ bin and lines of the various elements, 
obtained by setting all element abundances to zero except the one of interest,
calculated using the MEKAL code.
{\bf Top Panel}: the lines of O, Fe, Ne, Ni, Mg and Si used for the analysis
 in the 0.5-2 keV energy range.
{\bf Bottom Panel}: the lines of Si, S, Ar and Ca used for the analysis in the
2-4 keV energy range.
\label{lines}}
\end{figure}

In Figure~\ref{lines} we show the data of the 3$^{\prime}$-4$^{\prime}$ bin together
with the best fitting model calculated using the MEKAL code.
The model has been plotted nine times, each time all element abundances, 
except one, are set to zero. In this way the contribution of the various 
elements to the observed lines and line blends become apparent. In the energy 
band (0.5-4 keV) we have adopted for the spectral fitting, 
the abundance measurements based on
K-lines for all the elements except for Fe and Ni, 
for which the measure is based on L-lines.
The K-lines of O, Si, S, Ar and Ca are well 
isolated 
from other emission features and clearly separated from the continuum emission,
 which are the requirements for a robust measure of the equivalent width of the
lines and consequently of the abundances of these elements. 
The Fe-L lines are known to be problematic, because the atomic physics involved
is more complicated than K-shell transitions \citep{liedahl95}, but
from the very good signal of \emph{XMM} spectra and  
from the experience of \emph{ASCA} data  
\citep{mush96,hwang97,fuka98}
 we can conclude that the Fe-L determination is reliable.
Some of the stronger Fe-L lines due to Fe XXII and Fe XXIV are close to the
K-lines of Ne and Mg, respectively and blending can lead to 
errors in the Ne abundance and, to a smaller extent, to the Mg 
abundance \citep{liedahl95,mush96}. Also the Ni 
measure is difficult due to the possible confusion of its L-lines with the 
continuum and Fe-L blend.   

\subsection{Abundance profiles}

\begin{figure}
\epsscale{0.8}
\plotone{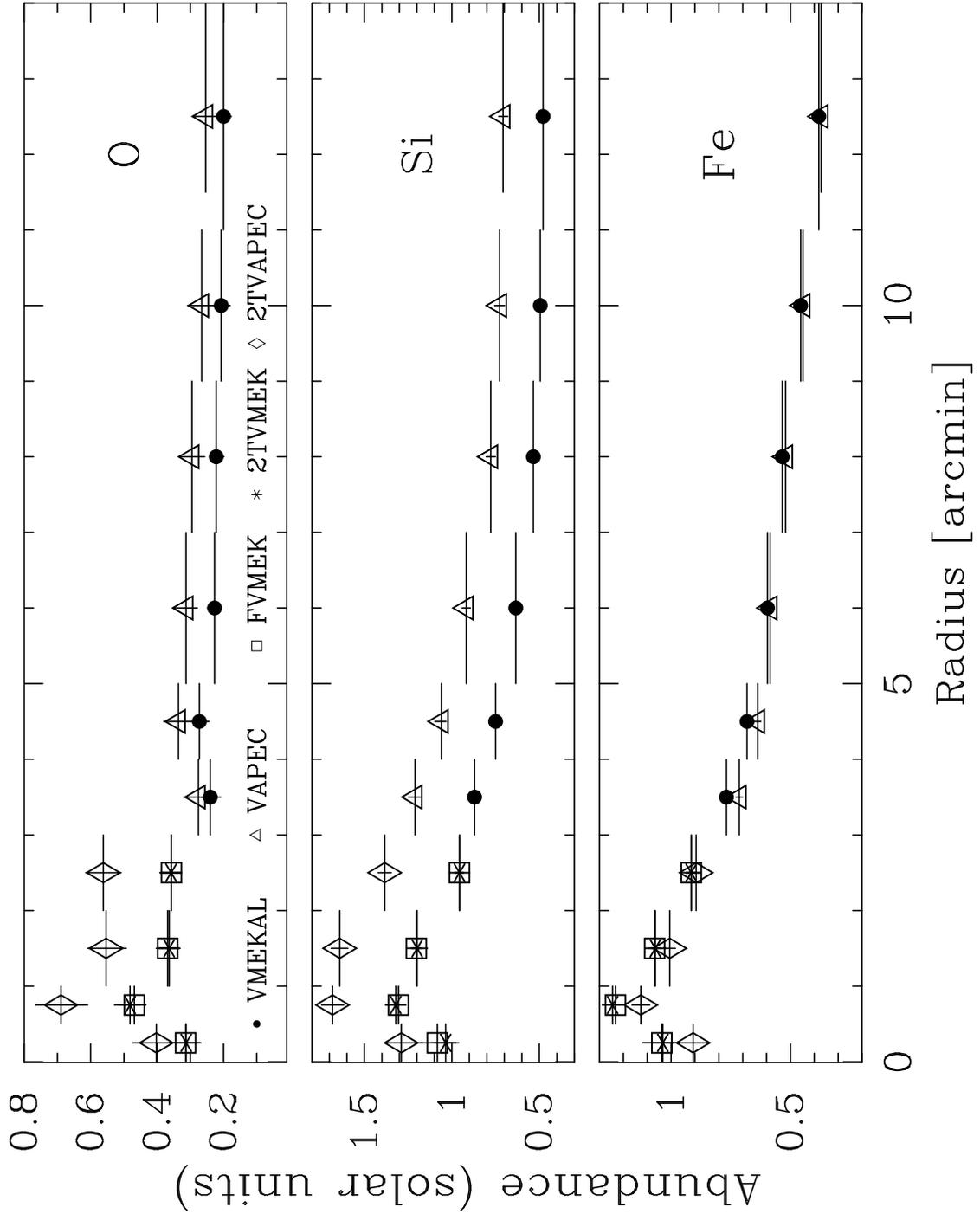}
\caption
{MOS abundance profiles for O, Si and Fe. Uncertainties are at the 68\% 
level for
 one interesting parameter($\Delta \chi^{2}\,=\,1$). Diamonds represent
measurements with 2T model using APEC code, asterisks those with 2T model 
using MEKAL code and squares those with fake multi-phase model using MEKAL code.
Full circles indicate measurements with 1T model using MEKAL code while empty
triangles those with 1T model using APEC code.
\label{z1}}
\end{figure}

\begin{figure}
\plotone{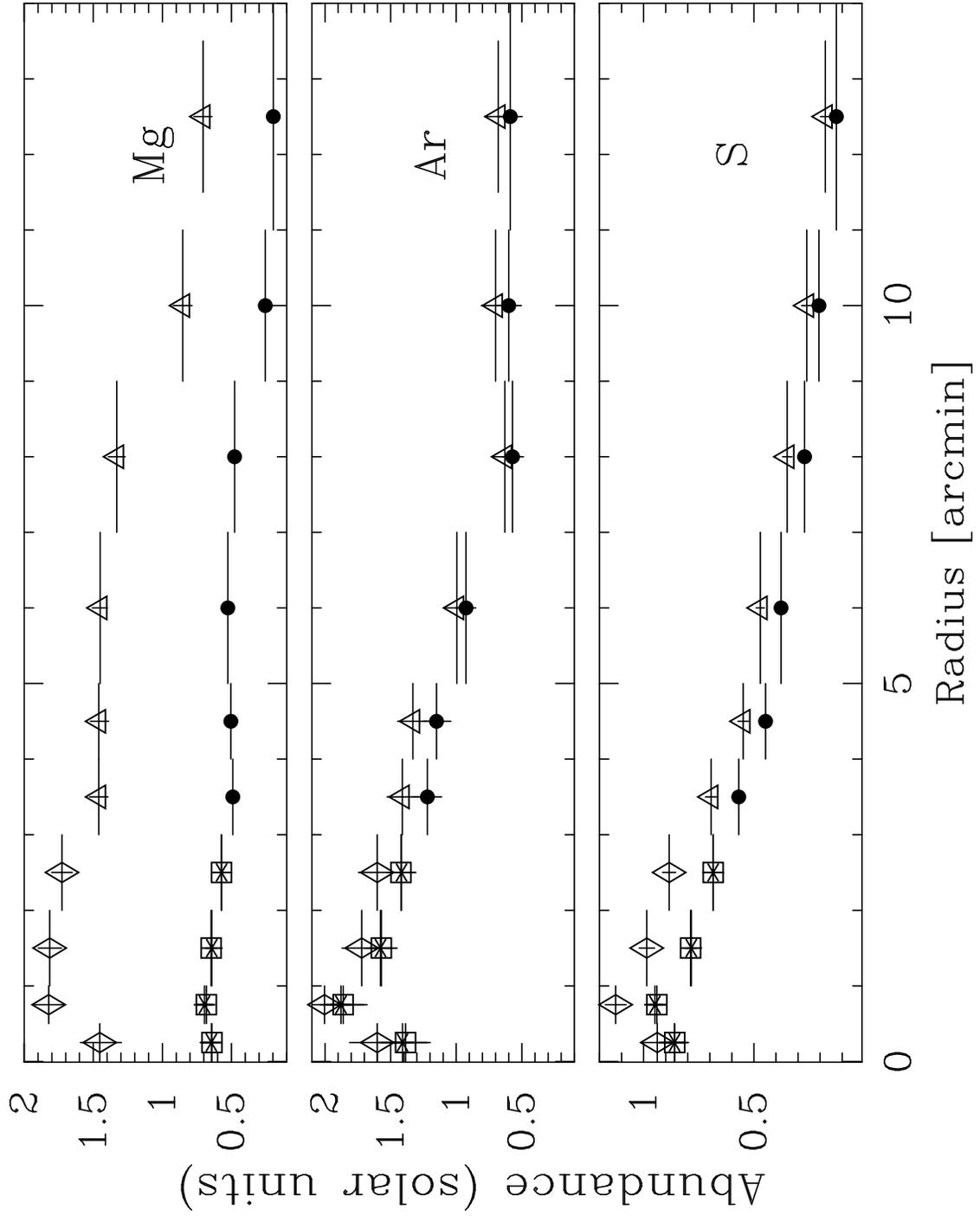}
\caption
{MOS abundance profiles for Mg, Ar and S. Symbols as in Figure~\ref{z1}.
\label{z2}}
\end{figure}

\begin{figure}
\epsscale{0.8}
\plotone{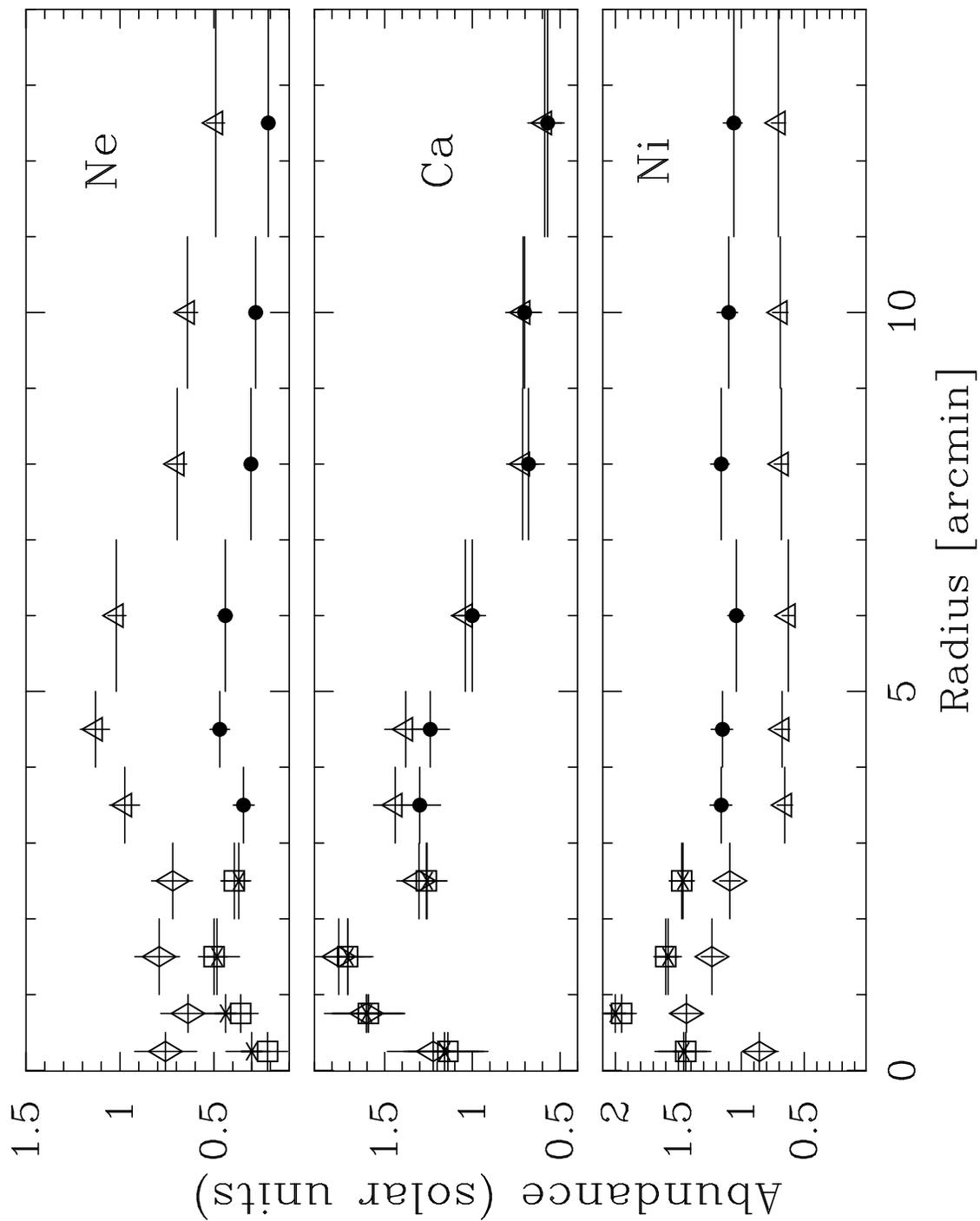}
\caption
{MOS abundance profiles for Ne, Ca and Ni. Symbols as in Figure~\ref{z2}.
\label{z3}}
\end{figure}

In Figure~\ref{z1} we report the MOS radial abundance  profiles for O 
(top panel), Si (middle panel) and Fe (bottom panel), in Figure~\ref{z2}
those for Mg (top panel), Ar (middle panel) and S 
(bottom panel), in Figure~\ref{z3} those for Ne (top panel), Ca 
(middle panel) and Ni (bottom panel).
We note that the measurements obtained using the two different plasma codes 
agree for what concerns Fe, Ar and Ca; they are somewhat 
different
for what concerns O, S and Ni and in complete 
disagreement for Mg and Ne.

\begin{figure}
\epsscale{0.8}
\plotone{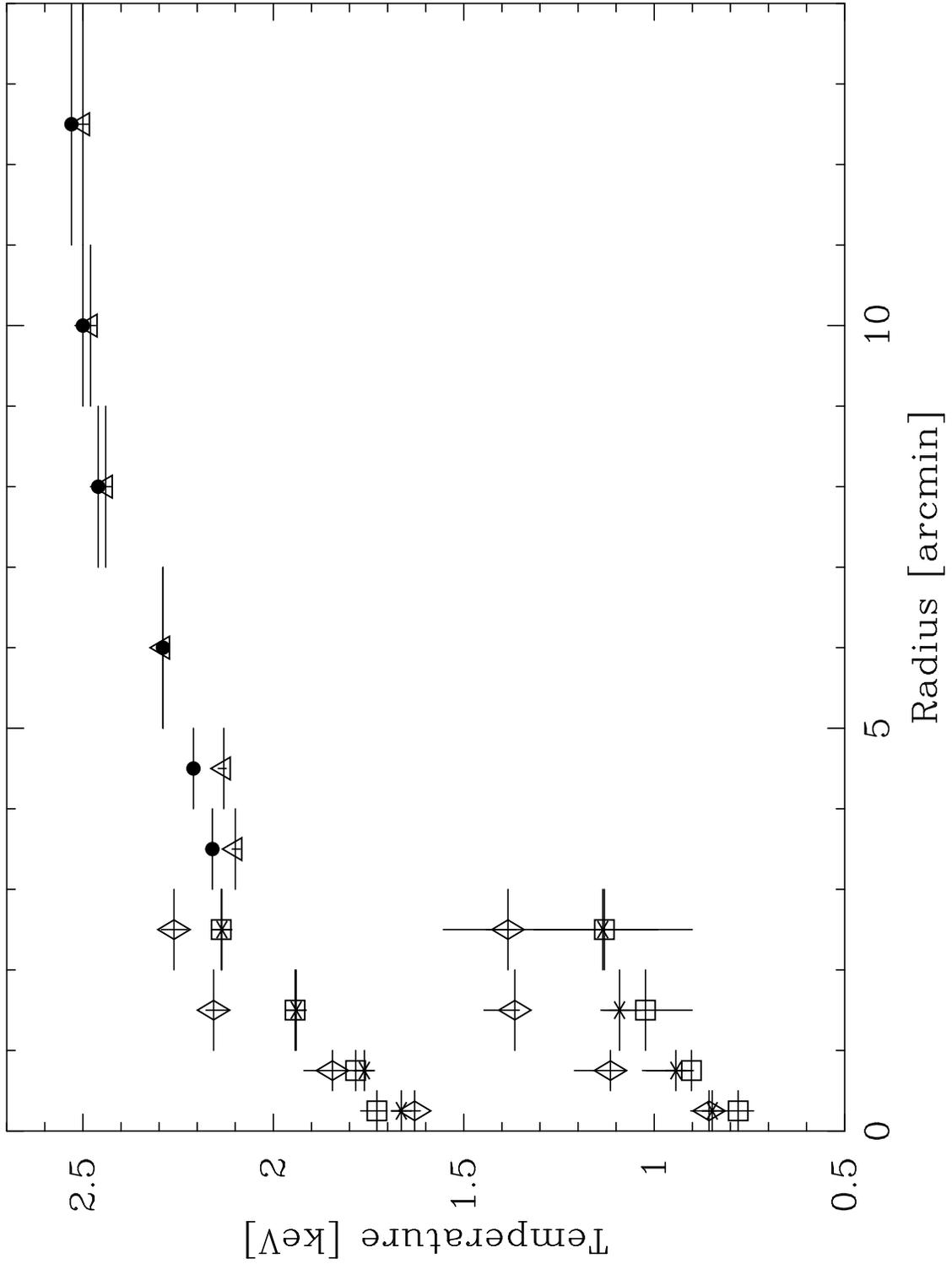}
\caption
{MOS temperature profile. Uncertainties are at the 68\% 
level for one interesting parameter($\Delta \chi^{2}\,=\,1$). 
Diamonds and asterisks
represent the two temperature values obtained with 2T models using APEC code 
and MEKAL code, respectively. Squares represent the value of the temperature 
of the vmekal component and the $T_{min}$ of the vmcflow component, for the
fake multi-phase model using MEKAL code. Full circles indicate temperature
 obtained with 1T model using MEKAL code while empty
triangles those with 1T model using APEC code.
\label{temperature}}
\end{figure}

The temperature profile obtained with the two codes is showed in Figure~\ref{temperature}: 
there are some differences in the inner regions, while in the outer isothermal
bins there is substantially agreement.

\begin{figure}
\vskip -2.0truecm
\epsscale{0.90}
\plotone{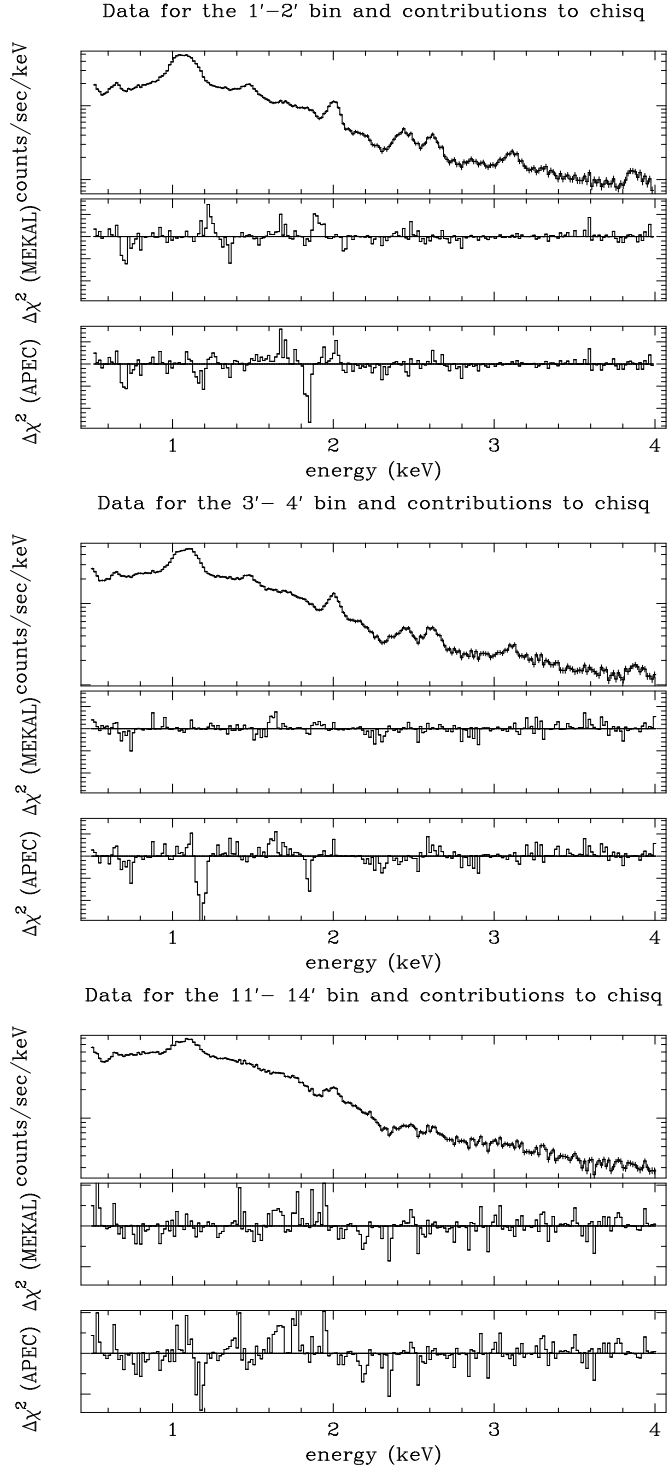}
\vskip -2.0truecm
\caption
{{\bf Top Panel}: data for the bin 
1$^{\prime}$-2$^{\prime}$ and contributions to $\chi^{2}$ for a 2T model with MEKAL code, the $\chi^{2}/d.o.f.$ of the fit is 381/217, and for a 2T model with APEC code, the $\chi^{2}/d.o.f.$ of the fit is 474/218. Contributions to $\chi^{2}$ are on the same scale  for direct comparison.
\newline
{\bf Middle Panel}: data for the bin 
3$^{\prime}$-4$^{\prime}$ and contributions to $\chi^{2}$ for a 1T model with MEKAL code, the $\chi^{2}/d.o.f.$ of the fit is 326/221, and for a 
1T model with APEC code, the $\chi^{2}/d.o.f.$ of the fit is 
546/222. Contributions to $\chi^{2}$ are on the same scale for direct 
comparison. 
\newline
{\bf Bottom Panel}: data for the bin 
11$^{\prime}$-14$^{\prime}$ and contributions to $\chi^{2}$ for a 1T model with MEKAL code, the $\chi^{2}/d.o.f.$ of the fit is 729/221, and for a 
1T model with APEC code, the $\chi^{2}/d.o.f.$ of the fit is 
886/222. Contributions to $\chi^{2}$ are on the same scale for direct 
comparison. 
\label{confapecmekal}}
\end{figure}

The models using the APEC code give a systematically worse description 
than the ones 
using MEKAL code: for the multi temperature models with APEC the
$\chi^{2}$ range from 403 to 498 for $\nu = 218$ (216 in the central bin 
due to the two additional degree of freedom of the power-law component), while
for MEKAL models (2T and fake multi-phase give the same results) the $\chi^{2}$
range from 323 to 382 for $\nu = 217$ (215 in the central bin 
due to the two additional degree of freedom of the power-law component); 
for single temperature models with APEC the $\chi^{2}$ range from 546 to 886
 for $\nu = 222$, while 
for MEKAL models the $\chi^{2}$ range from 326 to 731 for $\nu = 221$. In
Figure~\ref{confapecmekal} we compare the residuals in the form of $\Delta\chi^{2}$ between a 2T
 model using the MEKAL code and the APEC code for the inner bin 
1$^{\prime}$-2$^{\prime}$, as well as a 1T model using
 the MEKAL code and the APEC code for the ``isothermal'' 
bins 3$^{\prime}$-4$^{\prime}$ and 11$^{\prime}$-14$^{\prime}$. 
It's evident that in the  
external bins the differences in the fit between the two codes are due
to APEC over-prediction of the flux of Fe-L lines from high ionization states, 
considering the fact that the temperature obtained by the two codes are 
nearly coincident.
For the inner regions the differences between the two codes are further 
complicated by the different temperature range they find for the best fit. 
In general, where the temperature structure is very similar, as in the 
innermost bin, the
difference is as in the outer bins in the high energy part of the Fe-L blend,
 while where the temperature structure is different, as in the 
1$^{\prime}$-2$^{\prime}$ shown in Figure~\ref{confapecmekal}, the differences between the two 
codes are primary due to different estimates of the flux of He-like Si-K line.

We therefore choose as our best abundance profiles those obtained with 
a 2T vmekal fit for the central regions and with a 1T vmekal fit for the outer
regions.
In Table~\ref{abundances} we report the abundance profiles for O, Ne, Na, Mg, Al, Si, S, Ar, Ca, Fe and Ni obtained in this way.
Abundance gradients are clearly evident for Fe, Si, S,
Ar and Ca; O, Mg and Ni 
show evidence for an enhancement in the central regions while
only Ne is substantially flat.

\subsection{Comparison with Finoguenov et al. (2001) results}

We made a direct comparison of our results with the abundances 
and abundance ratios derived by \citet{fino01}. Their values are 
consistent within 
the errors with ours except for the oxygen abundance, which is roughly 
two times higher in Finoguenov's analysis.
To better understand the origin of this discrepancy 
we have extracted spectra from the same radial 
bins, 1$^{\prime}$-3$^{\prime}$ and
8$^{\prime}$-14$^{\prime}$, and fitted a single temperature model 
(vmekal) in the 
0.5-10 keV band, excising the 0.7-1.6 keV energy range, in order to avoid 
the dependence from the Fe L-shell peak, as done by \citet{fino01}. 
Our results for the O abundance, 
given in units of the solar values from \citet{anders89} to 
make a direct comparison with the results of Table 1 of \citet{fino01}, are
$0.32\pm0.03$ for the 1$^{\prime}$-3$^{\prime}$ bin and $0.20\pm0.02$ for the
8$^{\prime}$-14$^{\prime}$ bin, to be compared  with $0.535^{+0.019}_{-0.021}$ 
and $0.386^{+0.025}_{-0.021}$. Also interesting are the results for the Ni 
abundance: $2.55\pm0.33$ for the 1$^{\prime}$-3$^{\prime}$ bin and 
$2.34\pm0.26$ for the 8$^{\prime}$-14$^{\prime}$ bin, to be compared with 
$2.573^{+0.924}_{-0.918}$ and $0.800^{+1.732}_{-0.800}$.
We also note that a 2T modeling of the inner bin 1$^{\prime}$-3$^{\prime}$ 
gives a statistically better fit over a single temperature model 
( $\chi^{2}$/d.o.f of 919/440 respect to 1083/442, using the 0.5-10 keV band)
breaking down the assumption of near isothermality. 

We also cross compare the K- and L-shell results for Ni and Fe in our analysis,
 performing the spectral fits for all the radial bins in the 0.5-10 keV band, 
but excising the 0.7-1.6 keV energy range. The derived abundances for the two
elements are consistent within $1\sigma$, although the K-shell Ni abundance 
is 20$\%$ higher than the L-shell abundance in the bins fitted with a single
temperature model. It should be borne in mind that particularly for the 
outer bins the K-shell Ni abundance measure is very sensitive to 
the background estimate.   

\subsection{Abundance ratios and SNIa Fe mass fraction}

\begin{figure}
\plotone{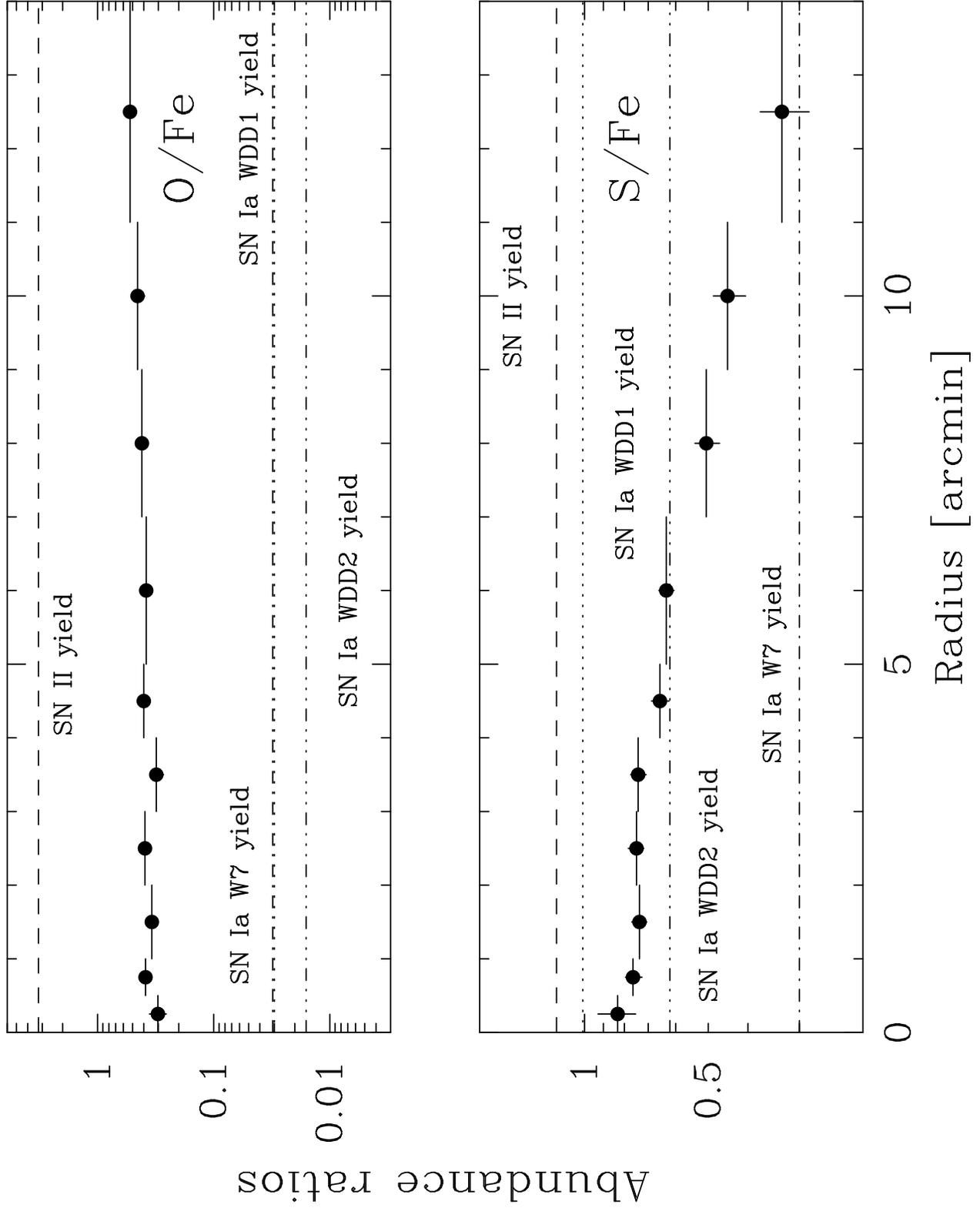}
\caption
{MOS abundance ratio profiles for O/Fe and S/Fe. Also showed are the 
abundance ratios predicted by SNe models taken by \citet{nomoto97}: the dashed
line refers to the SNII model, the dash-dotted line to the W7 SNIa model, the
dotted line to the WDD1 SNIa model and the three dotted-dashed line to the 
WDD2 SNIa model.
\label{ratio1}}
\end{figure}

\begin{figure}
\epsscale{0.8}
\plotone{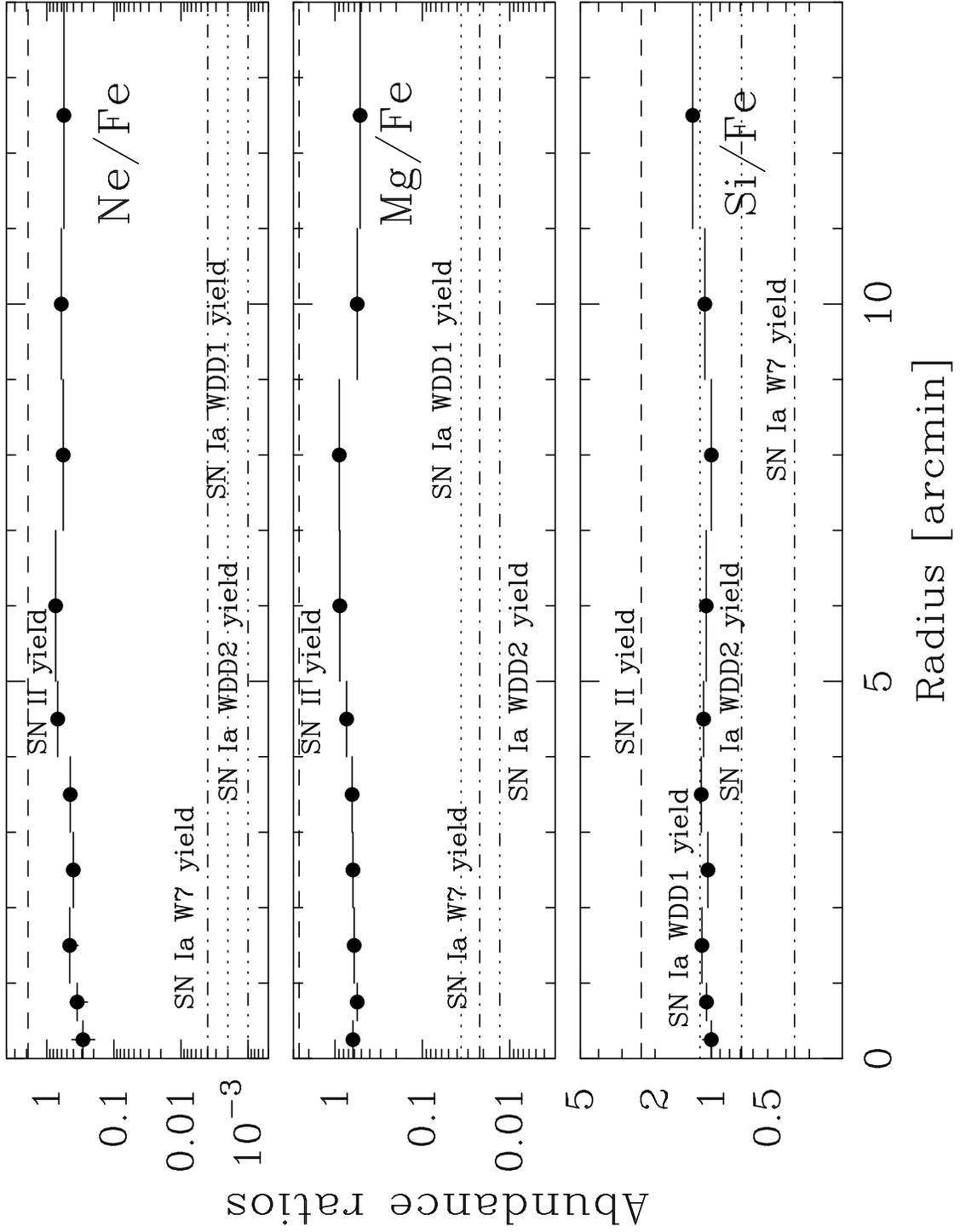}
\caption
{Same as Figure~\ref{ratio1} but for ratio profiles for Ne/Fe, Mg/Fe and Si/Fe 
\label{ratio2}}
\end{figure}

\begin{figure}
\epsscale{0.8}
\plotone{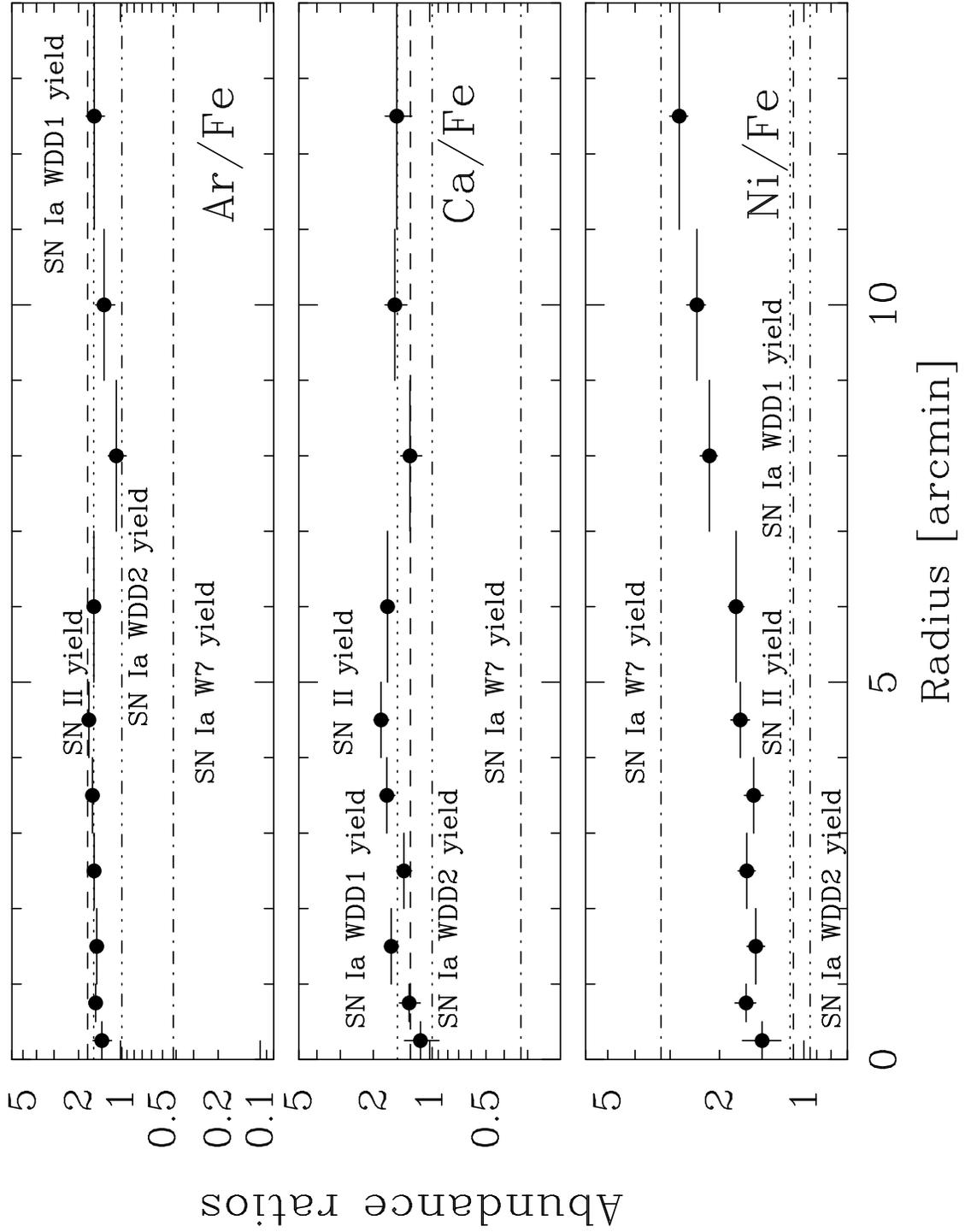}
\caption
{Same as Figure~\ref{ratio1} but for ratio profiles for Ar/Fe, Ca/Fe and Ni/Fe.
\label{ratio3}}
\end{figure}

\begin{figure}
\epsscale{0.8}
\plotone{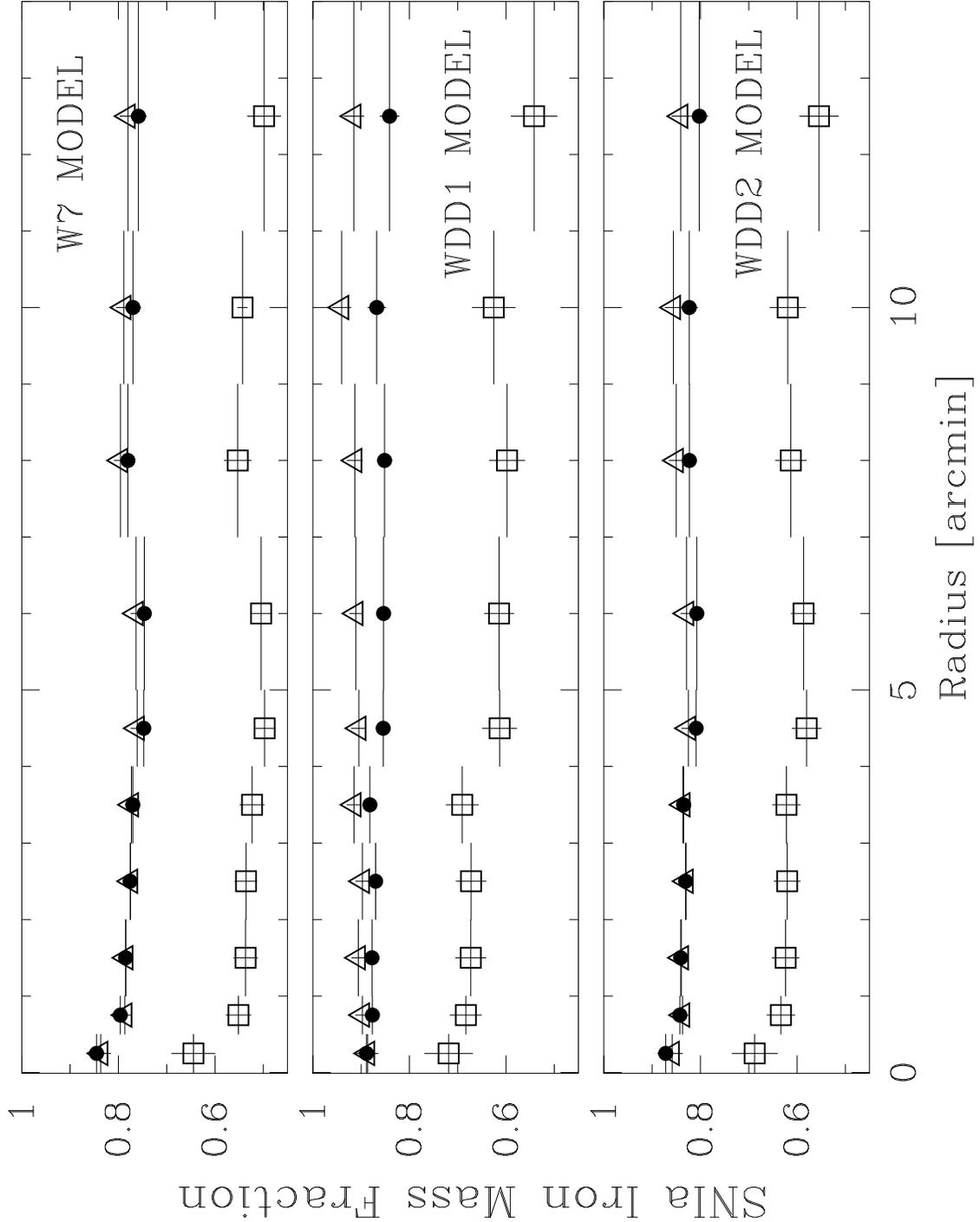}
\caption
{{\bf Top Panel} SNIa Fe Mass Fraction obtained by a simultaneous fit of the eight abundance 
ratios of Table 2, using the W7 model for SNIa and different yields for SNII: 
the ones by Nomoto et al.(1997) (circles), 
the upper end of the range indicated by Gibson et al.(1997) (triangles) and 
the lower end of the range (squares). Errors are at 68\% 
confidence limit(1$\sigma$).
\newline
{\bf Middle Panel} Same as the top panel but using WDD1 model for SNIa.
\newline
{\bf Bottom Panel} Same as the top panel but using WDD2 model for SNIa.
\label{sniamassfractio}}
\end{figure}

From the abundance measurements we obtain the abundance ratios between all
the elements relative to Fe, normalized to the
solar value. They are shown in 
Figure~\ref{ratio1}, Figure~\ref{ratio2}, Figure~\ref{ratio3} and in Table~\ref{ratios}, together with the abundance ratios obtained 
by models of supernovae taken by \citet{nomoto97} and rescaled to the 
solar abundances reported in \citet{grevesse98}.
We use those abundance ratios to estimate the relative contributions of
SNIa and SNII to the metal enrichment of the intra-cluster gas. Such estimates
are complicated by uncertainties both in the observations and in the 
theoretical yields.
Our approach is to use the complete set of ratios trying to find the best fit
of the function

\begin{equation}
\label{fit}
\Big(\frac{X/Fe}{X_{\odot}/Fe_{\odot}}\Big)_{observed}\,=\,f\,\Big(\frac{X/Fe}{X_{\odot}/Fe_{\odot}}\Big)_{SNIa}+(1-f)\,\Big(\frac{X/Fe}{X_{\odot}/Fe_{\odot}}\Big)_{SNII}
\end{equation}
 
where $\Big(\frac{X/Fe}{X_{\odot}/Fe_{\odot}}\Big)_{observed}$ is the measured abundance ratio of the $X$ element to $Fe$, given in solar units, 
$\Big(\frac{X/Fe}{X_{\odot}/Fe_{\odot}}\Big)_{SNIa}$ and
$\Big(\frac{X/Fe}{X_{\odot}/Fe_{\odot}}\Big)_{SNII}$ are the theoretical abundance ratio by the two types of supernovae, also given in solar units and $f$ is directly the SNIa Fe mass fraction. The result of the 
simultaneous fit of the eight ratios is presented as circles in Figure~\ref{sniamassfractio}, 
using the SNII model by \citet{nomoto97} and W7, WDD1 and WDD2 models for SNIa respectively for the three panels.
Due to the large uncertainties in the yields for the SNII model, the results 
are strongly SNII model dependent. For comparison we use the range of SNII 
yields calculated by \citet{gibson97}, also listed in Table~\ref{ratios}, which 
involve only ratios for O, Ne, Mg, Si and S and as before finding the best fit for eq.(\ref{fit}). The results are
 shown as
 squares for the lower end of the range and triangles for the upper 
part of the range.

\begin{figure}
\vskip -2.0truecm
\epsscale{1.0}
\plotone{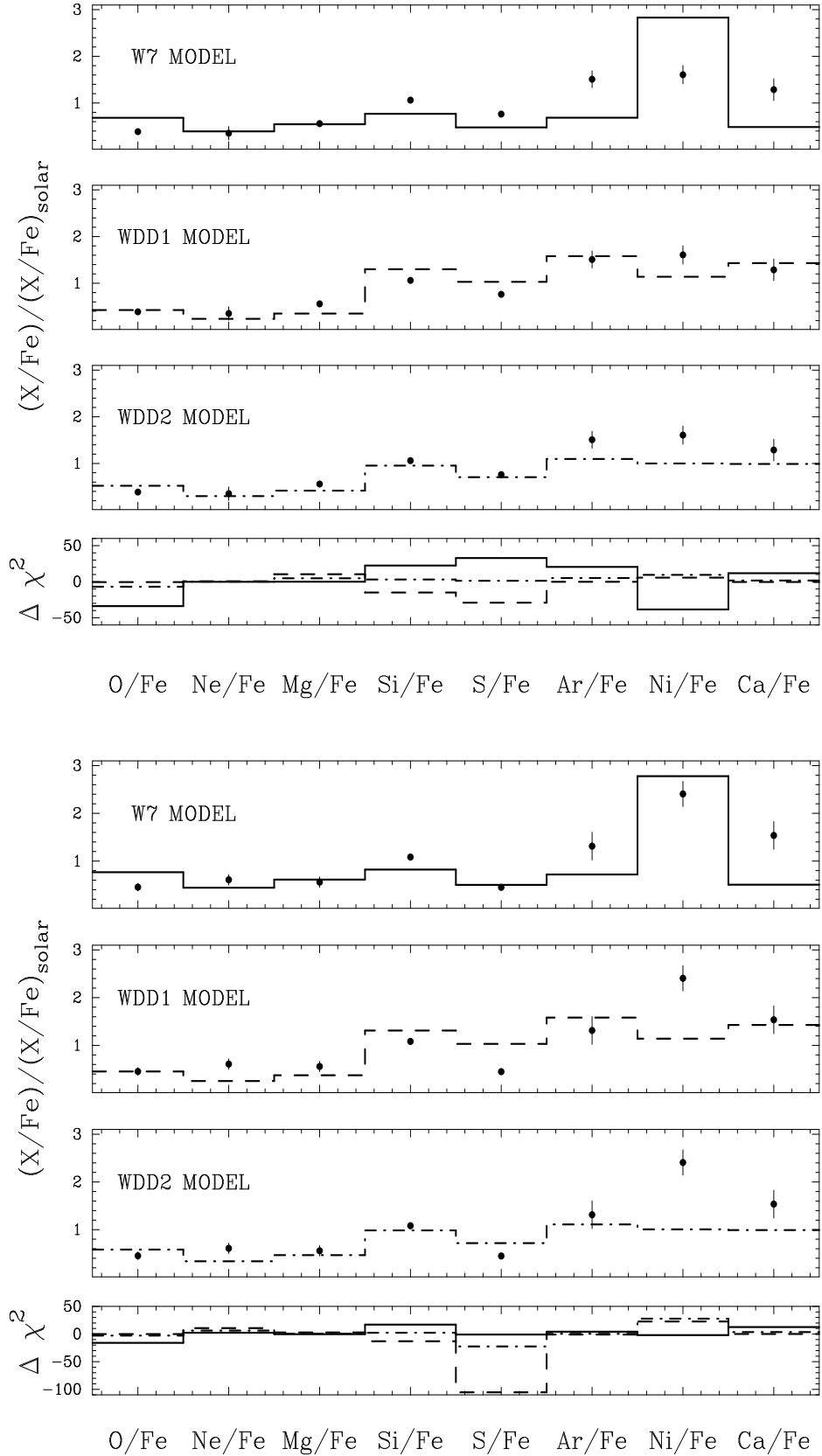}
\vskip -1.0truecm
\caption
{{\bf Top Panel} Abundance ratios for the 0.5$^{\prime}$-1$^{\prime}$ bin, best fits for the Nomoto SNIa models, in combination with the Nomoto SNII model, and contributions to $\chi^{2}$ for the various models. The solid line refers to the W7 model, the dashed line to the WDD1 model and the dot-dashed line to the WDD2 model.
\newline
{\bf Bottom Panel} Same as the top panel but for the 11$^{\prime}$-14$^{\prime}$ bin.
\label{fitsnia}}
\end{figure}

The best fits are obtained in the inner bins with a combination of the WDD2 
model for SNIa and the Nomoto model for SNII, with reduced $\chi^{2}$ which 
ranges from 2 in the inner bin up to 10 in the 4$^{\prime}$-5$^{\prime}$ bin.
In the outer bins the fit is slightly better (reduced $\chi^{2}$ of 8-10 
instead of 12-13) with a combination of the W7 model for SNIa and the Nomoto
model for SNII. This is shown in Figure~\ref{fitsnia} where the fits with the 
three SNIa models together with the Nomoto SNII model are reported for the 
0.5$^{\prime}$-1$^{\prime}$ bin and for the 11$^{\prime}$-14$^{\prime}$ bin.
It's clear from the inspection of the residuals in terms of $\Delta\chi^{2}$ 
that the combination of W7 model and Nomoto SNII model fails in the inner bins 
because it predicts a higher Ni/Fe 
and O/Fe ratio, compared to the delayed detonation models.
The situation is the opposite for the outer bins where a higher Ni/Fe 
and a lower S/Fe favors the W7 model. However the preference for the W7
model in combination with the Nomoto SNII model in the outer bins is 
strongly dependent on the Ni/Fe ratio: if
we exclude it from the fit the WDD2 model provides the better fit to the data.

\section{Discussion}

The model emerging from the \emph{ASCA} and \emph{BeppoSAX} data for 
the explanation of 
abundance gradients in galaxy clusters was that of a homogeneous enrichment by
SNII, the main source of $\alpha$ elements, maybe in the form of strong 
galactic winds in the proto-cluster phase and
the central increase in the heavy element distribution due to an enhanced 
contribution by SNIa, strongly related to the presence of a cD galaxy 
\citep{fuka98,dupke00b,fino00,grandi01,makishima01}.
As a textbook example we can consider the case of A496 observed by 
\emph{XMM-Newton} 
\citep{tamura01}. The O-Ne-Mg abundance is radially constant over the 
cluster, while the excess of heavy elements as Fe, Ar, Ca and Ni in the core is
consistent with the assumption that the metal excess is solely produced by
SNIa in the cD galaxy. The crucial ratio for the discrimination of the 
enrichment by the two types of supernovae, O/Fe, is then decreasing towards 
the center.

The \emph{XMM} results for M87/Virgo question this picture. They confirm 
and improve the accuracy of the measure of heavy elements gradients 
previously found by
\emph{ASCA} and \emph{BeppoSAX}, but they also show a a statistically significant enhancement
of  $\alpha$ elements O and Mg in the core. If we consider the 
inner 9 arcmin the ratio O/Fe is constant with $\chi^{2}=6.9$ for 7 d.of. and 
adding a linear component does not improve the fit  
($\chi^{2}=5.4$ for 6 d.o.f). These facts points toward an increase in 
contribution also of SNII, since O is basically produced only by this kind of
supernovae. Although there is little or no evidence of current star formation 
in the core of M87, the O excess could  be related to a recent past episode of
 star formation triggered
by the passage of the radio jet, as we see in cD galaxies with a radio source (A1795 cD: van Breugel et al. 1984 and A2597 cD: Koekemoer et al. 1999), nearby (Cen A: Graham 1998) and
distant radio galaxies \citep{vanbreugel85,vanbreugel93,bicknell00} 
(for a comprehensive discussion see McNamara 1999).
\newline 
To put the above idea quantitatively, waiting for a true deprojection of our 
data, we use previous \emph{ROSAT} estimate of 
the deprojected electron density in the center of 
the Virgo cluster \citep{nulsen95} to calculate the excess mass of oxygen. To 
estimate the excess abundance we fit the inner bins, where we see the 
stronger increase in the O abundance, with a constant obtaining an abundance
of 0.32, while for the outer bins we obtain an abundance of 0.21 so the excess
is 0.11. Then we estimate the oxygen mass, $M_{\rm{O}}$ to be $M_{\rm{O}} = A_{\rm{O}}\,y_{\rm{O},\odot}\,Z_{\rm{O}}^{excess}\,M_{H}$, where $M_{H} = 0.82\,n_{e}\,\frac{4\pi}{3}\,(R_{out}-R_{in})^{3}$, $R_{in}$ and $R_{out}$ being the bounding radii in kpc of the bins,  $A_{\rm{O}}= 16$ and $y_{\rm{O},\odot} = 6.76 \times 10^{-5}$ from \citet{grevesse98}. We obtain a rough estimate of $10^{7}\,\rm{M_{\odot}}$ which, assuming 
that 
about $80\,\rm{M_{\odot}}$ of star formation are required to generate a SNII 
\citep{thomas90} and Nomoto SNII oxygen yield, requires a cumulative star 
formation of $4-5 \times 10^{7}\,\rm{M_{\odot}}$. This star formation is in 
agreement with a burst mode of star formation ($\lesssim 10^{7}$ yr) at 
rates of 
$\sim 10-40\,\rm{M_{\odot}\,yr^{-1}}$, as it is observed in the CD 
galaxies of A1795 and A2597 \citep{mcnamara99}.

For what concern heavy element gradients and the contribution of SNIa, M87 data
suggests an agreement with delayed detonation models (in particular for the 
inner bins), as stressed by 
\citet{fino01}, in contrast with the preference of W7 model set by the high Ni/Fe ratios found by \citet{dupke00a}. 
This is particular evident if we consider the S/Fe 
ratio in Figure~\ref{ratio1}. If we consider the set of theoretical values for 
SNII and W7 SNIa models the behavior of these ratio would indicate an 
increasing contribution by SNIa going \emph{outward} to the center. 
We recover the
correct behavior if we choose the WDD1 yield and we reduce the S SNII yield of
\citet{nomoto97} by a factor of two to three, as was already indicated by 
\emph{ASCA} data \citep{dupke00a}. We caution however 
that this is a substantial
contribution larger than that allowed by the SNII models choosed by 
\citet{gibson97}.
The use of delayed detonation model for SNIa could also explain the over 
abundance of S and also Si (respect to the W7 model) found by \citet{tamura01} for the core of A496.
\newline
With increasing radius the W7 model gives a better fit to the data respect to 
delayed detonation models. This fact indicates a SNIa abundance pattern 
change with radius and could be taken as an independent X-ray 
confirmation of the conclusions of \citet{hatano00} on the optical 
spectroscopic diversity of SNIa, as suggested by \citet{fino01}.
However we stress that the preference of W7 over WDD models in the outer bins 
is entirely due 
to the Ni/Fe ratio which could be affected by systematic uncertainties, as 
discussed in section 5.1 and 5.3.

In Figure~\ref{sniamassfractio}, we show  the relative importance of
SNIa, measured by the Fe mass fraction provided by this kind of supernovae.
 This is 
substantially constant through the 14 arcmin analyzed. This fraction is 
considerable and ranges between the 50\% and 80\% and it depends only slightly 
on the SNIa model used. Instead the uncertainties involved in using different SNII models are large and a 
definitive answer cannot be reached until further convergence of SNII models
is achieved. 

\section{Summary}
We have performed a spatially resolved measurement of the element abundances in
M87, the Virgo cluster cD galaxy. The main conclusion of our work are:

\begin{itemize}
\item the APEC code gives a systematically worse description than the MEKAL 
code in modeling M87 spectra;
\item we confirm the increase of Fe and other heavy elements towards the core 
indicating that the SNIa contribution increases;
\item the increase in O abundance and a constant O/Fe ratio in the inner 9 
arcmin indicates an increase also in SNII ejecta possibly from star-burst in the
recent past;
\item Si/Fe and S/Fe profiles favor WDD models over W7, also requiring substantial reduction of the SNII yield of S;
\item the indication of a change of the SNIa abundance pattern, provided by a preference of the W7 model over delayed detonation models in the outer bins, is entirely due to the Ni/Fe ratio. Since the Ni measurement is difficult and uncertain, this indication should be taken with some caution.
\end{itemize}

\acknowledgments
S. Ettori and S. Ghizzardi are thanked for useful discussions and suggestions.
We thank the referee for several suggestions that improved the presentation of this
work.
 This work is based on observations obtained with \emph{XMM-Newton}, an ESA science mission with instruments and contributions directly funded by ESA Member States and the USA (NASA).



\clearpage


\clearpage

\clearpage
\epsscale{0.8}

\clearpage

\clearpage

\clearpage

\clearpage

\clearpage

\clearpage

\clearpage

\clearpage

\clearpage


\begin{deluxetable}{cccccccccc}
\tablecolumns{10}
\tablewidth{0pt}
\tablecaption{Abundance profiles obtained by a 2T vmekal modeling for the
central regions and a 1T vmekal modeling for the outer regions. All abundances
referred to solar units as given in \citet{grevesse98}. All errors 
quoted are at the 68\% level for one interesting parameter ($\Delta\chi^{2} = 1$)
\label{abundances}}
\tablehead{
\colhead{Bin}  & \colhead{O} & \colhead{Ne} & \colhead{Mg} & \colhead{Si} & \colhead{S} & \colhead{Ar} & \colhead{Ca} & \colhead{Fe} & \colhead{Ni}} 
\startdata
0$^{\prime}$-0.5$^{\prime}$ & $0.31^{+0.05}_{-0.04}$ & $0.30^{+0.14}_{-0.10}$ &
$0.64^{+0.08}_{-0.05}$ & $1.03^{+0.08}_{-0.07}$ & $0.86^{+0.07}_{-0.06}$ &
$1.41^{+0.19}_{-0.19}$ & $1.16^{+0.23}_{-0.22}$ & $1.03^{+0.08}_{-0.07}$ &
$1.46^{+0.23}_{-0.18}$ \\ 
0.5$^{\prime}$-1$^{\prime}$ & $0.48^{+0.04}_{-0.05}$ & $0.44^{+0.12}_{-0.13}$ &
$0.69^{+0.05}_{-0.06}$ & $1.32^{+0.04}_{-0.04}$ & $0.95^{+0.03}_{-0.04}$ &
$1.88^{+0.12}_{-0.18}$ & $1.60^{+0.20}_{-0.20}$ & $1.24^{+0.03}_{-0.04}$ &
$2.00^{+0.19}_{-0.14}$ \\
1$^{\prime}$-2$^{\prime}$   & $0.36^{+0.04}_{-0.03}$ & $0.48^{+0.10}_{-0.12}$ &
$0.64^{+0.04}_{-0.04}$ & $1.20^{+0.03}_{-0.03}$ & $0.78^{+0.03}_{-0.02}$ &
$1.58^{+0.12}_{-0.13}$ & $1.71^{+0.14}_{-0.14}$ & $1.07^{+0.03}_{-0.02}$ &
$1.58^{+0.11}_{-0.11}$ \\
2$^{\prime}$-3$^{\prime}$   & $0.36^{+0.03}_{-0.03}$ & $0.37^{+0.09}_{-0.06}$ &
$0.57^{+0.03}_{-0.04}$ & $0.96^{+0.03}_{-0.03}$ & $0.68^{+0.03}_{-0.02}$ &
$1.42^{+0.11}_{-0.11}$ & $1.26^{+0.11}_{-0.11}$ & $0.92^{+0.02}_{-0.02}$ &
$1.46^{+0.10}_{-0.09}$ \\
3$^{\prime}$-4$^{\prime}$   & $0.24^{+0.03}_{-0.03}$ & $0.34^{+0.05}_{-0.06}$ &
$0.49^{+0.03}_{-0.04}$ & $0.87^{+0.02}_{-0.02}$ & $0.57^{+0.02}_{-0.02}$ &
$1.22^{+0.11}_{-0.11}$ & $1.30^{+0.11}_{-0.12}$ & $0.77^{+0.01}_{-0.02}$ &
$1.16^{+0.09}_{-0.08}$ \\
4$^{\prime}$-5$^{\prime}$   & $0.27^{+0.03}_{-0.03}$ & $0.47^{+0.05}_{-0.05}$ &
$0.50^{+0.03}_{-0.03}$ & $0.75^{+0.02}_{-0.02}$ & $0.45^{+0.02}_{-0.02}$ &
$1.15^{+0.10}_{-0.11}$ & $1.24^{+0.11}_{-0.11}$ & $0.68^{+0.01}_{-0.01}$ &
$1.15^{+0.09}_{-0.08}$ \\
5$^{\prime}$-7$^{\prime}$   & $0.23^{+0.02}_{-0.02}$ & $0.44^{+0.04}_{-0.04}$ &
$0.53^{+0.03}_{-0.03}$ & $0.63^{+0.02}_{-0.01}$ & $0.38^{+0.01}_{-0.01}$ &
$0.93^{+0.07}_{-0.07}$ & $1.00^{+0.08}_{-0.07}$ & $0.60^{+0.01}_{-0.01}$ &
$1.04^{+0.07}_{-0.06}$ \\
7$^{\prime}$-9$^{\prime}$   & $0.22^{+0.02}_{-0.02}$ & $0.30^{+0.03}_{-0.03}$ &
$0.48^{+0.03}_{-0.03}$ & $0.53^{+0.02}_{-0.02}$ & $0.27^{+0.01}_{-0.01}$ &
$0.57^{+0.08}_{-0.08}$ & $0.68^{+0.08}_{-0.09}$ & $0.53^{+0.02}_{-0.02}$ &
$1.16^{+0.08}_{-0.06}$ \\
9$^{\prime}$-11$^{\prime}$  & $0.21^{+0.02}_{-0.03}$ & $0.28^{+0.03}_{-0.03}$ &
$0.25^{+0.03}_{-0.03}$ & $0.49^{+0.02}_{-0.02}$ & $0.20^{+0.02}_{-0.02}$ &
$0.60^{+0.09}_{-0.10}$ & $0.70^{+0.09}_{-0.10}$ & $0.46^{+0.01}_{-0.01}$ &
$1.10^{+0.09}_{-0.07}$ \\
11$^{\prime}$-14$^{\prime}$ & $0.20^{+0.02}_{-0.02}$ & $0.21^{+0.03}_{-0.03}$ &
$0.20^{+0.03}_{-0.03}$ & $0.48^{+0.02}_{-0.01}$ & $0.13^{+0.02}_{-0.02}$ &
$0.59^{+0.08}_{-0.09}$ & $0.57^{+0.09}_{-0.09}$ & $0.38^{+0.01}_{-0.01}$ &
$1.06^{+0.08}_{-0.07}$ \\
\enddata
\end{deluxetable}

\clearpage

\begin{table}
\caption{Abundance ratios from our results and from SNe models. The abundances 
relative to Fe normalized to the solar value, 
(X/Fe)/($\rm{X}_{\odot}/\rm{Fe}_{\odot}$), where X and Fe are number density
of the element and Fe, are shown. The errors associated with the observed 
abundance ratios are the propagated $1\sigma$ errors.
\label{ratios}}
\begin{center}
\vspace{1.0cm}
\begin{tabular}{|c|c|c|c|c|c|c|c|c|c|}
\hline
  & O & Ne & Mg & Si & S & Ar & Ca & Ni \cr
0$^{\prime}$-0.5$^{\prime}$ & $0.30^{+0.06}_{-0.04}$ & $0.29^{+0.13}_{-0.10}$ &
$0.62^{+0.09}_{-0.06}$ & $1.00^{+0.12}_{-0.10}$ & $0.83^{+0.10}_{-0.08}$ &
$1.36^{+0.21}_{-0.20}$ & $1.12^{+0.24}_{-0.23}$ & $1.41^{+0.25}_{-0.20}$ \cr
0.5$^{\prime}$-1$^{\prime}$ & $0.39^{+0.03}_{-0.04}$ & $0.35^{+0.10}_{-0.10}$ &
$0.56^{+0.04}_{-0.05}$ & $1.06^{+0.04}_{-0.05}$ & $0.76^{+0.03}_{-0.04}$ &
$1.51^{+0.10}_{-0.15}$ & $1.29^{+0.16}_{-0.17}$ & $1.61^{+0.15}_{-0.12}$ \cr
1$^{\prime}$-2$^{\prime}$   & $0.34^{+0.04}_{-0.03}$ & $0.45^{+0.09}_{-0.11}$ &
$0.60^{+0.04}_{-0.04}$ & $1.12^{+0.04}_{-0.05}$ & $0.73^{+0.03}_{-0.03}$ &
$1.48^{+0.12}_{-0.12}$ & $1.60^{+0.14}_{-0.14}$ & $1.48^{+0.11}_{-0.10}$ \cr
2$^{\prime}$-3$^{\prime}$   & $0.39^{+0.04}_{-0.03}$ & $0.40^{+0.10}_{-0.07}$ &
$0.62^{+0.04}_{-0.04}$ & $1.04^{+0.04}_{-0.04}$ & $0.75^{+0.03}_{-0.03}$ &
$1.55^{+0.12}_{-0.12}$ & $1.37^{+0.13}_{-0.13}$ & $1.60^{+0.12}_{-0.11}$ \cr
3$^{\prime}$-4$^{\prime}$   & $0.31^{+0.04}_{-0.04}$ & $0.45^{+0.07}_{-0.07}$ &
$0.64^{+0.04}_{-0.06}$ & $1.13^{+0.04}_{-0.04}$ & $0.74^{+0.03}_{-0.03}$ &
$1.59^{+0.15}_{-0.14}$ & $1.69^{+0.15}_{-0.16}$ & $1.51^{+0.12}_{-0.11}$ \cr
4$^{\prime}$-5$^{\prime}$   & $0.40^{+0.04}_{-0.04}$ & $0.69^{+0.08}_{-0.08}$ &
$0.74^{+0.05}_{-0.05}$ & $1.10^{+0.04}_{-0.04}$ & $0.65^{+0.03}_{-0.03}$ &
$1.69^{+0.15}_{-0.16}$ & $1.82^{+0.16}_{-0.16}$ & $1.69^{+0.14}_{-0.12}$ \cr
5$^{\prime}$-7$^{\prime}$   & $0.38^{+0.04}_{-0.04}$ & $0.74^{+0.07}_{-0.06}$ &
$0.88^{+0.05}_{-0.05}$ & $1.06^{+0.03}_{-0.03}$ & $0.63^{+0.03}_{-0.03}$ &
$1.55^{+0.13}_{-0.13}$ & $1.68^{+0.13}_{-0.13}$ & $1.74^{+0.12}_{-0.11}$ \cr
7$^{\prime}$-9$^{\prime}$   & $0.41^{+0.04}_{-0.04}$ & $0.57^{+0.06}_{-0.06}$ &
$0.89^{+0.07}_{-0.06}$ & $1.00^{+0.05}_{-0.04}$ & $0.50^{+0.03}_{-0.04}$ &
$1.07^{+0.15}_{-0.16}$ & $1.27^{+0.16}_{-0.17}$ & $2.17^{+0.17}_{-0.14}$ \cr
9$^{\prime}$-11$^{\prime}$  & $0.45^{+0.05}_{-0.06}$ & $0.61^{+0.08}_{-0.08}$ &
$0.56^{+0.08}_{-0.07}$ & $1.08^{+0.04}_{-0.04}$ & $0.45^{+0.04}_{-0.04}$ &
$1.31^{+0.19}_{-0.21}$ & $1.54^{+0.20}_{-0.21}$ & $2.41^{+0.21}_{-0.16}$ \cr
11$^{\prime}$-14$^{\prime}$ & $0.52^{+0.05}_{-0.06}$ & $0.55^{+0.07}_{-0.08}$ &
$0.52^{+0.09}_{-0.09}$ & $1.26^{+0.05}_{-0.04}$ & $0.33^{+0.04}_{-0.05}$ &
$1.54^{+0.22}_{-0.24}$ & $1.50^{+0.23}_{-0.25}$ & $2.78^{+0.22}_{-0.18}$ \cr

\hline

W7$^{a}$ & 0.031 & 0.004 & 0.022 & 0.36 & 0.30 & 0.42 & 0.33 & 3.23 \cr
\hline
WDD1$^{a}$ & 0.030 & 0.002 & 0.036 & 1.15 & 1.01 & 1.56 & 1.48 & 1.12 \cr
\hline
WDD2$^{a}$ & 0.016 & 0.001 & 0.013 & 0.69 & 0.62 & 0.98 & 0.97 & 0.95 \cr
\hline 
SNII$^{b}$ & 3.23  & 1.90  & 2.58  & 2.37 & 1.17 & 1.72 & 1.09 & 1.27 \cr
\hline
SNII$^{c}$ & 1.25-2.80 & 0.96-1.93 & 0.96-1.89 & 1.81-2.54 & 0.85-2.08 & - & - & - \cr
\hline
\end{tabular}
\end{center}
\vskip 1.0truecm
$^{a}$ Different models of SNIa taken by \citet{nomoto97}.

$^{b}$ Yields of SNII taken by \citet{nomoto97}.

$^{c}$ \citet{gibson97} who choose a representative sample of SNII yields 
in literature.

\end{table}

\end{document}